**Sustainable Metal-Organic Framework Water Harvesters in the Artificial Intelligence Era**

Reid A. Coyle[1], Shyam Chand Pal[1], Peter Walther[1], Saeun Park[1], Bin Feng[1,2], Zhiling Zheng[1,2]*

[1] Department of Chemistry, Washington University, St. Louis, MO 63130, United States.
[2] Institute of Materials Science & Engineering, Washington University, St. Louis, MO 63130, United States.

*E-mail: z.z@wustl.edu

## Abstract

Metal–organic frameworks (MOFs) are excellent candidates for water harvesting due to their tunable pore environments, which can be precisely engineered to capture and release water in arid conditions. Integrating artificial intelligence (AI) into MOF discovery can further accelerate the design of high-performance sorbents by identifying structural features that enhance atmospheric water harvesting (AWH), stability, and cycling efficiency. In this Perspective, we examine key MOF design principles, including cooperative adsorption, operational relative humidity (RH), uptake capacity, hysteresis, and scalability. We highlight recent design advancements such as multivariate strategies and long-arm linker extension, and examine how these principles tune pore capacity and hydrophilicity, while preserving stability and crystallinity. Furthermore, we discuss how AI, large language models (LLMs), and data mining can accelerate the discovery process through predictive synthesis, inverse design, and elucidating synthesis-structure-property relationships for the next generation of MOF water harvesters.

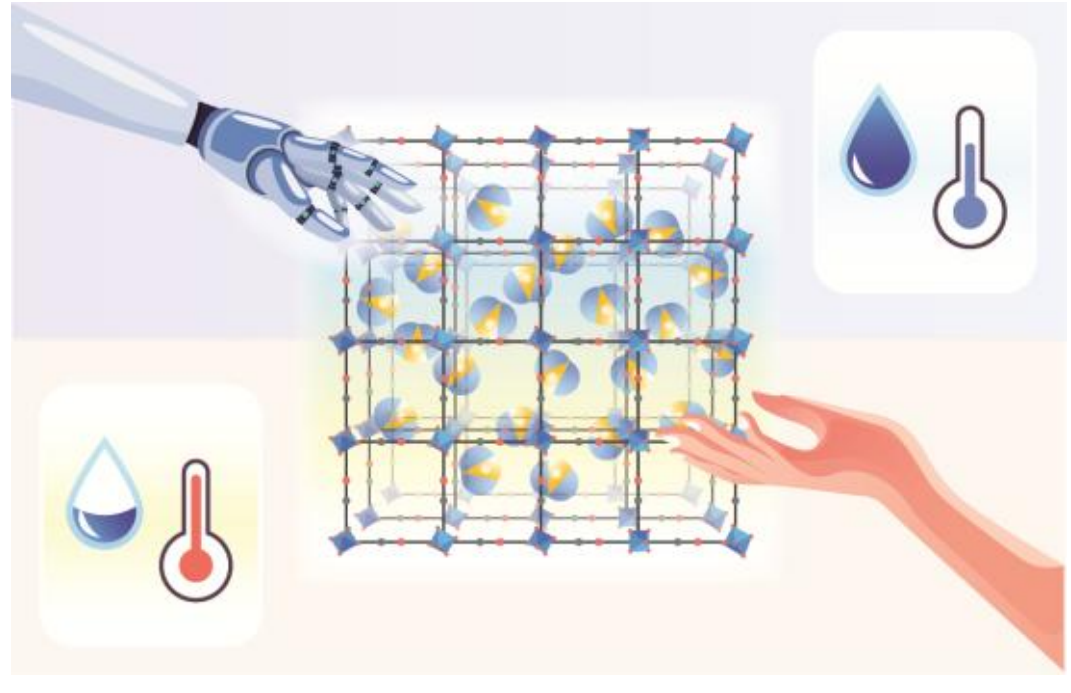

## Water Harvesting MOF

In light of escalating water scarcity worldwide, particularly in arid and semi-arid regions, the quest for clean and sustainable water sources has never been more urgent.[1–4] This urgency is accentuated by the widespread presence of large-scale, water-intensive sectors, such as data centers,[5–7] agricultural and food processing,[8,9] and thermoelectric power generation.[10–12] These demands often overlap with desert-like ecosystems across all continents, affecting over 100 countries (**Fig. 1**) and impacting the livelihoods of a rapidly growing population reliant on dependable drinking water sources.[1,13,14] Atmospheric water, abundant and universally accessible, emerges as a promising solution, but its effective harnessing hinges on the development of porous and hygroscopic sorbents.[15–18]

Recently, MOFs stand at the vanguard of this endeavor due to their remarkable ability to be engineered at the atomic level and have been successfully employed in field tests in real desert environments.[19–23] The key to this success is the ability to intricately shape the behavior of water molecules, influencing crucial aspects such as productivity, operational conditions, kinetics, recyclability, and energy efficiency.[21,24,25] Notably, the past ten years have witnessed a surge in research interest in this area, with contributions from over five hundred institutions worldwide (**Fig. 1**), underscoring the global recognition of the importance of this technology in addressing water scarcity. In this Perspective Article, we highlight the recent development of the design strategies of MOFs tailored for AWH, focusing on the pivotal factors influencing the AWH process and particularly how the unique interior features of MOFs influence cooperative water isotherms. Bridging experimental insights with AI approaches, we aim to illuminate the relationship between water isotherms and the design and properties of MOF materials. This exploration not only aims to deepen our understanding of MOF-based water harvesting systems but also seeks to address a fundamental question: How can we develop better MOF water harvesters?

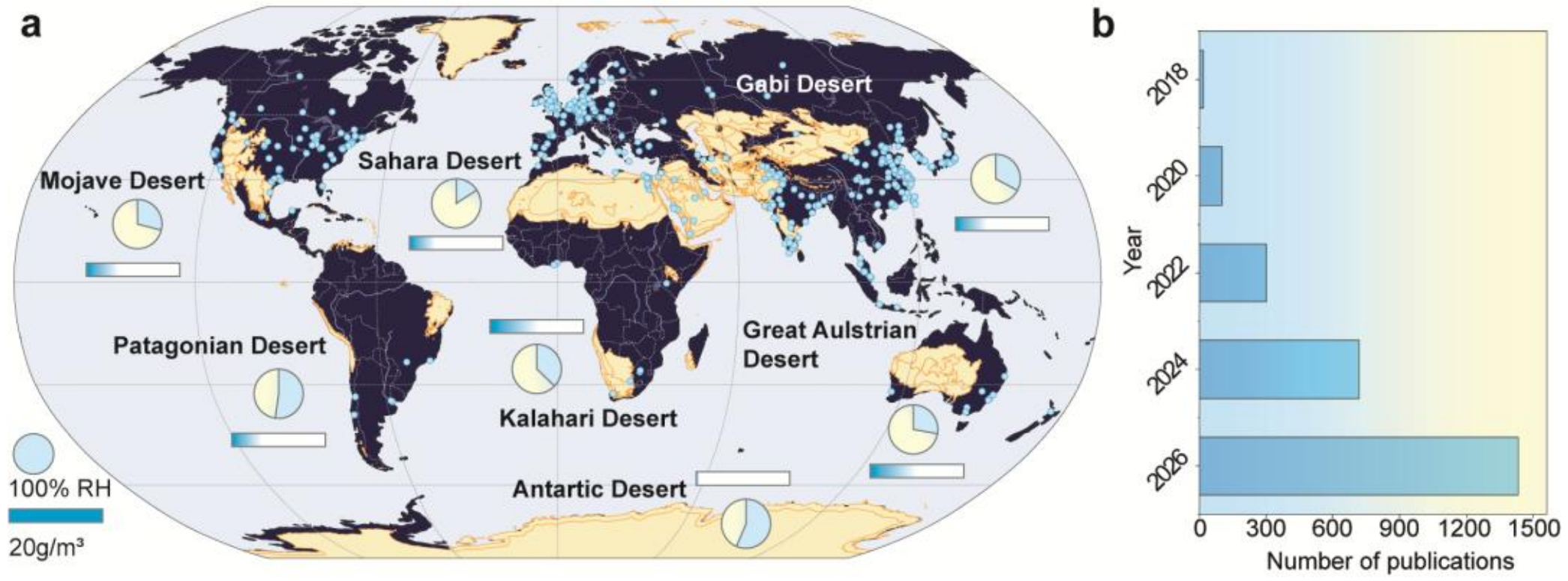

**Figure 1 | a,** Global distribution of desert ecosystems and research interest in MOF-based water harvesting. The map highlights both hot and cold desert ecosystems[26] across 92 countries, depicted in bright yellow. The bright blue dots indicate research institutes engaged in the development of MOFs for water harvesting. Pie charts depict RH, and adjacent bars show absolute humidity, with a full bar representing 20g $m^{-3}$. **b,** Total accumulated publications with keywords "metal–organic frameworks" and "machine learning" in Web of Science as of May 20, 2026.

**Key considerations for water harvesting MOF**

During the past two decades, extensive research has been done on the sorption isotherms of various gases (such as $N_2$, $H_2$, $CH_4$, $CO_2$, $H_2O$, etc.), owing to their high surface area and porosity which provide incredible storage space and adsorption sites for these gas molecules.[27–32] Although the concept of MOF capturing water vapor was established in early 2000s, the practical application of MOFs in atmospheric water harvesting from desert air, along with the development of MOF-based water harvester devices, did not gain significant traction until the mid-2010s.[33] This shift marked a transition from the primary focus on characterizing water isotherms in MOFs to a more targeted approach towards designing MOFs for efficient water harvesting from desert air.

Following this paradigm shift, the focus in the field has progressively shifted towards the deliberate design of MOFs at the atomic level to shape their isotherm profiles (**Fig. 2**). Herein, we focus on the interpreting features of water isotherms in MOFs and discuss the key considerations that define an efficient water harvesting MOF.

*Shape of the Cooperative Isotherm.* A cornerstone of this field is the recognition of the correlation between the cooperative isotherm shape and the efficiency for water harvesting.[32,34,35] The appearance of step-shaped water sorption isotherms in MOFs is indicative of a unique process: the initial formation of seeding water molecules at adsorptive sites followed by the subsequent arrangement into ordered, hydrogen-bonded networks within the crystalline structure during the pore-filling process.[24,34,36] At the isotherm level, this phenomenon is manifested as a sudden increase in water uptake at a specific RH, resulting in an S-shape isotherm (**Fig. 2**). Importantly, contrasting with continuous uptake isotherms, the step-shaped profile is particularly preferred for water harvesting applications, as it allows for facile water uptake and release through relatively small changes in temperature or pressure gradients. This attribute positions MOFs with suitable seeding water adsorption sites and pore environment as energetically favorable sorbents, especially at low RH levels.

*Operational RH Range.* Efficient water harvesting from the arid desert air typically necessitates MOFs with a low operational humidity range (e.g. 5—30%).[14] In the case of MOFs exhibiting a single-step isotherm, the determination of the operational RH is straightforward: it is near the position of the isotherm's step—below this RH cut-off, the MOF interacts minimally with water molecules, whereas above it, the sorbent is fully loaded with a large quantity of water. The position of this step is intrinsically linked to the hydrophilicity of the pore environment, which is influenced by both the metal secondary building units (SBUs) and the organic linkers on the backbone. Consequently, a step positioned towards the lower end of the RH on the isotherm indicates a MOF's proficiency in extracting water from more arid environments.

On the other hand, MOFs characterized by multistep water isotherms, often resulting from a mesoporous environment or a physical mixture of two or more MOFs, present a more complex scenario.[37–39] These MOFs, due to their multistep adsorption behavior, are typically less favored for water harvesting applications. The complexity arises from having to choose between operating with partial capacity, which reduces the total productivity accumulatively, or demanding higher energy costs associated with the need for greater changes in temperature or pressure to release adsorbed water (**Fig. 2**).

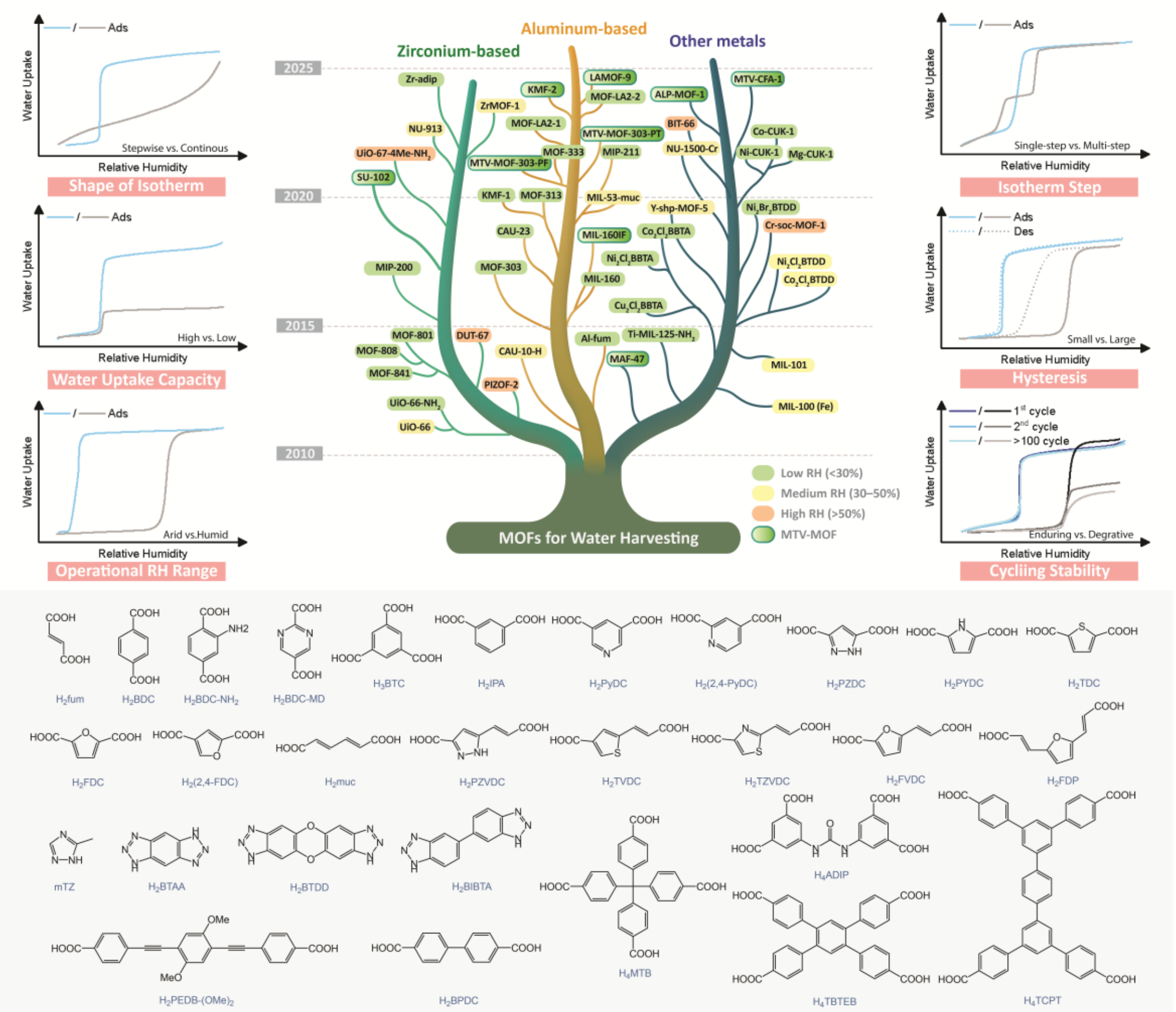

**Figure 2** | Development of water harvesting MOFs and analysis of their water isotherms. The tree diagram in the top panel showcases some of the representative MOFs reported for water harvesting from 2010 to 2026.[40–49] Six isotherm plots juxtaposed show the shape of water absorption under desirable (blue curves) and undesirable (grey curves) characteristics. The bottom panel shows the variety of MOF linkers utilized in constructing water harvesting MOFs.

*Water Uptake Capacity.* Water uptake capacity, reflected by the vertical axis of the isotherm, is a key metric for assessing MOF sorbents for water harvesting. The uptake capacity of MOFs, which varies depending on the metal used, can range from 0.3 g/g to over 1.0 g/g (>100% wt.), which has been addressed in several comprehensive summaries.[15,50–52] Notably, for any specific MOF, this uptake value is directly related to the pore volume and crystallinity of the compound. If the synthesis of the MOF is not fully optimized to achieve maximum porosity, the resultant water uptake capacity will be compromised.[25,53]

Additionally, we note that simply considering the water uptake of MOFs prepared at a small synthesis scale can be misleading. Assessing whether the retention of water uptake capacity is maintained at a larger synthesis scale (e.g., >100g or even at the kilogram level) and how many cycles can be performed at scale is crucial, as an essential prerequisite for the future broader application of MOF-based AWH technology is the scalability of MOF processing.[21,54–62] Ideally, MOFs synthesized at a large scale should exhibit the same water uptake characteristics as those produced on a smaller scale.

*Hysteresis.* The presence of a hysteresis loop should be avoided, as additional energy is required to dissociate water clusters during the regeneration process.[28,32,52] Typically, this phenomenon arises from irreversible capillary condensation in mesoporous materials with pore diameters larger than 2 nm. Furthermore, structural changes induced by functional groups and the flexibility of the framework, as well as less crystalline compounds resulting from unoptimized synthesis, can also contribute to the occurrence of the hysteresis loop.[63,64]

**Human-guided and AI-assisted design of new water harvesting MOFs**

The future of water harvesting MOFs lies in atomic level structural engineering. By combining insights from experimentation with AI-driven inverse design and predictive synthesis, researchers can infer synthesis–structure–property relationships and shape water isotherms for practical desert water harvesting.[65] Specifically, by shaping the isotherm curve horizontally (to adjust uptake) and vertically (to set the RH cut-off), it is conceptually possible to create on-demand water harvesting materials that are not only energy-efficient, but also capable of maximizing productivity under varying operational conditions. The concept of harvesting water anytime and anywhere moves closer to reality with MOFs that can be structurally programmed to adapt to diverse real-world conditions. As emerging design strategies converge with AI and computational methods, a new generation of water-harvesting MOFs with targeted isotherm profiles and climate-specific performance is coming into focus.

*Multivariate Strategy.* In the domain of reticular chemistry, the multivariate (MTV) approach offers a strategic pathway for synthesizing MOFs by introducing compositional heterogeneity within a singular framework, thereby creating variability across unit cells without altering their topology.[66–70] By adjusting the compositional ratios of metals and organic linkers, it is possible to precisely tailor the hydrophilic properties of the internal surfaces, enabling a shift in the RH cut-off for water capture.

Taking MOF-303 as a foundational framework, by substituting a portion of the pyrazole linker with an isoreticular but compositionally distinct linker, the hydrophilicity of the pore environment can be modulated. This concept gave rise to two groups of new MOFs, including MTV-MOF-303-PF series and MTV-MOF-303-PT series, which incorporate furan- and thiophane-based linkers, respectively.[24,71] Furthermore, post-synthetic metal exchange offers an additional strategy for tuning water-harvesting MOFs; the combination of two ligands will improve hydrolytic stability and facilitate water sorption.[72] These modifications have yielded operational improvements, including a decrease in regeneration temperatures by 10°C to 16°C and a reduction in desorption enthalpy of 5 kJ $mol^{-1}$. In addition, the MTV approach is also applicable to metal SBUs to shift of the isotherm in a similar manner.[47,73–75] This innovative approach has been increasingly adopted in recent times, leading to the development of diverse and efficient water-harvesting MOF families.[47,76–79]

*Long-arm Linker Extension Strategy.* The enhancement of water uptake in MOFs is fundamentally linked to the expansion of pore volume, which is achievable by linker extension. One commonly used approach to increase the linker length is to introduce additional phenyl rings.[30,80–82] However, it is not without trade-offs, as it often leads to a more hydrophobic environment and reduced hydrolytic stability.[83–86] Computational tools such as data mining can help identify promising linker candidates, repurpose existing frameworks, and accelerate the discovery of robust MOFs.[87–91]

Leveraging our understanding of MOF-303's base structure, computational tools can simulate water uptake post-expansion, and assume retention of the preferred isoreticular structure. Notably, the insights show that use of vinyl groups as 'long-arm' linkers has demonstrated enhanced pore volume and water uptake while maintaining hydrolytic stability, leading to the successful discovery of MOF-LA2-1.[92] Similarly, another example of linker extension from the shorter fumaric acid to the longer muconic acid was also recently reported.[93] In both cases, this close connection between molecular design and performance has enabled nearly a 50% increase in water harvesting, with lower desorption temperatures, while still maintaining the ability to harvest water at low humidity.

*Integrating MTV and Linker Extension Strategies.* Considering that there are two strategies capable of shifting the water isotherm in both directions, it was demonstrated that a series of long-arm MOFs, referred to as LAMOF-1 to LAMOF-10,[25] can be prepared to shift isotherms both horizontally (via the MTV approach) and vertically (via the long-arm approach), thereby demonstrating varied operational RH, water uptake capacities, and regeneration temperatures. By combining LLM-identified, synthetically accessible building blocks with DFT-elucidated adsorption sites, this approach provides a theoretical framework that complements experimental findings and guides the design of MOF synthesis with targeted water-harvesting properties. AI-guided predictive synthesis can further optimize crystallinity and pore accessibility, enabling continuous improvement of water uptake capacity.[94–98] This versatility allows the materials to meet diverse demands across different environmental conditions.

## MOF water harvester device

A good sorbent represents only the initial step in the journey towards practical water harvesting. The subsequent pivotal phase involves integrating these sorbents into functional devices, which fall into two primary categories: active and passive systems. Active MOF water harvesters, which operate through multicyclic processes, leverage external electricity, often derived from solar panels. In contrast, passive MOF water harvesters depend solely on ambient sunlight and natural cooling, typically functioning in a monocyclic manner (**Fig. 3**). For both cases, the fundamental steps are the same: (i) the sorbent is exposed to the air, and water vapor is adsorbed and concentrated in the sorbent; (ii) the sorbent is then enclosed in a closed space, where, through temperature or pressure changes, water is released, condensed, and collected as drinkable water.[22,99] A 2019 study in Mojave Desert highlighted a compact device leveraging solar power to drive multiple cycles of atmospheric water collection throughout the day.[13] Subsequently, in a 2022 study, a portable, hand-carried device was tested in Death Valley, one of the most extreme environments on Earth, during the summer. Remarkably, this device operates solely on thermal energy derived from sunlight, without any additional energy inputs.[14] Conceptually, LLMs and other AI driven approaches can provide guidance in scaling up this technology.[100] Furthermore, adapting the system to a continuous AWH process through the incorporation of a rotational cylinder is a promising avenue.[101] Insights gleaned from field tests have been pivotal in emphasizing the significance of shaping water isotherm for optimal operation of MOF-based devices and can be categorized as follows:

(1) Designing MOFs requires balancing hydrophobic and hydrophilic pore environments, as low RH cutoffs increase recycling energies, whereas high RH cutoffs lower capacity in arid conditions.

(2) Long-term stability is vital for the practical use of these devices and makes the MOF water harvester cost-effective.

(3) Factors such as airflow, heat, and mass transfer are vital for achieving peak efficiency under real-world operational conditions.

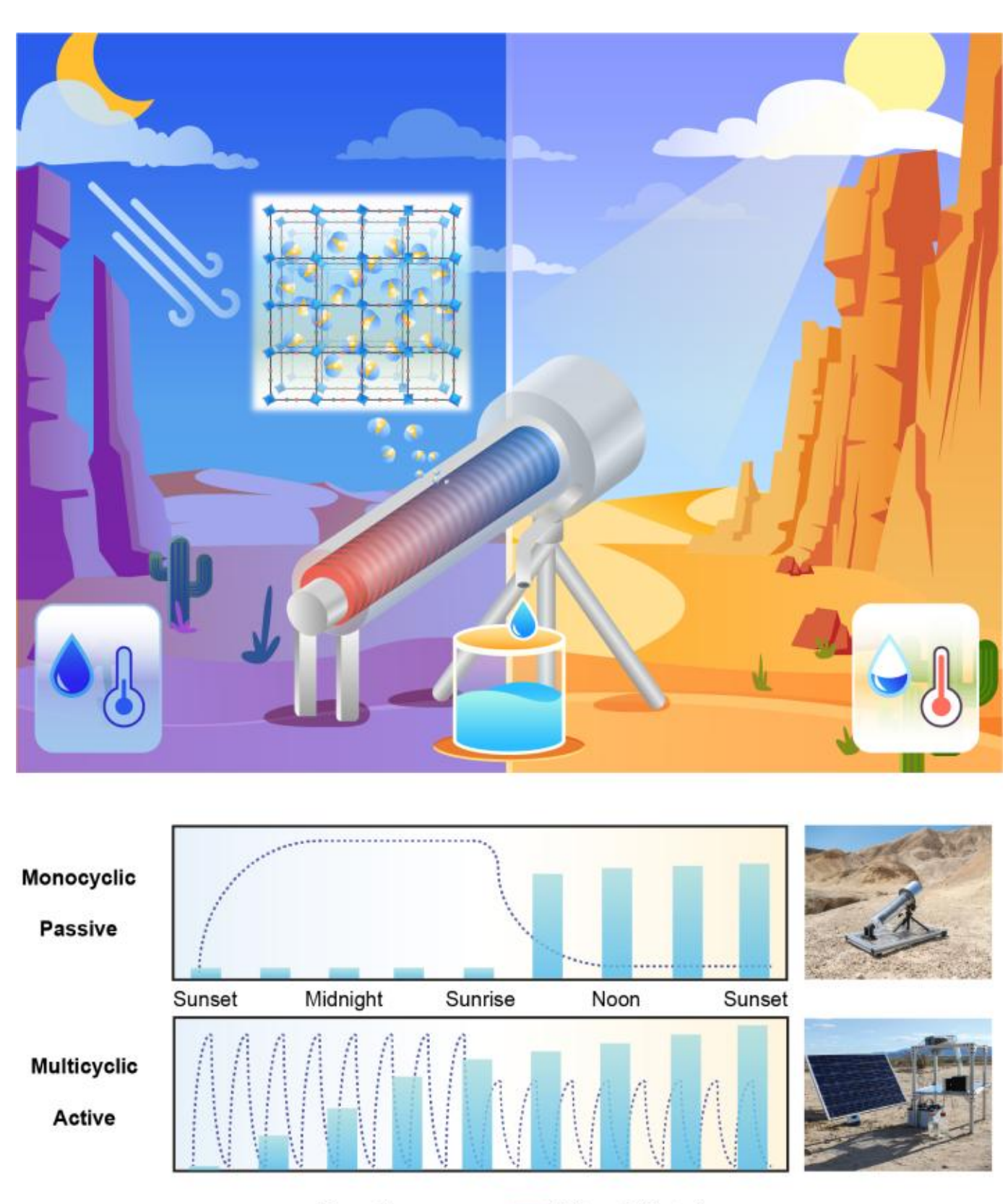

**Figure 3** | Field test of MOF water harvesters in desert. The top panel illustrates a passive device loaded with MOF-303. The bottom panel presents a comparative analysis between monocyclic and multicycle devices.

**Future Outlook**

We have highlighted recent advancements in the design strategies of MOFs for water harvesting from the air, and it has evolved from preliminary stages, where not a single device had been tested, to a vibrant research landscape in which hundreds of research institutions are now actively engaged in developing sorbents with tailored behaviors and devising a myriad of device architectures. A pivotal element propelling this advancement is the integration of AI and computational tools across atomic, molecular and device scales that has accelerated the discovery of MOF sorbents and the optimization of AWH systems. Looking forward, we posit that the integration of emerging machine learning models and LLMs will further accelerate the development process through rational design and performance improvement.[102–106] This synergy will continue to refine sorbent and device performance beyond the current benchmarks, heralding a new era of MOF water harvesting technologies for sustainable water production in water-scarce regions.

**Competing Financial Interests**

The authors declare no conflict of interests.

**Acknowledgments**

We thank Luz K. Molina for discussions on desert ecoregions and guidance on the geospatial datasets. Z.Z. gratefully acknowledges financial support from the EQT Foundation through the Breakthrough Science Grant (No. 059994) and extends sincere appreciation for their support.

**Author Biographies**

**Reid A. Coyle** is currently a Ph.D. student in the Department of Chemistry at Washington University in St. Louis. His research focuses on reticular chemistry design and the integration of artificial intelligence tools.

**Shyam Chand Pal** is a postdoctoral researcher in the Department of Chemistry at Washington University in St. Louis. His work focuses on the synthesis of metal–organic frameworks and covalent organic frameworks for applications in gas storage and separation.

**Peter Walther** is currently a Ph.D. student in the Department of Chemistry at Washington University in St. Louis. His research involves literature-based data mining and machine learning approaches for predicting reticular synthesis outcomes.

**Saeun Park** is currently a Ph.D. student in the Department of Chemistry at Washington University in St. Louis. Her research is focused on the synthesis of metal–organic frameworks for gas storage and separation applications.

**Bin Feng** is a Ph.D. student in the Department of Chemistry at Washington University in St. Louis. His research focuses on developing artificial intelligence agents to accelerate materials discovery.

**Zhiling Zheng** is currently an assistant professor in the Department of Chemistry at the Washington University in St. Louis. His research focuses on data-driven predictive synthesis of crystalline porous materials through large language models and high-throughput experimentation.